\documentclass[onecolumn,preprint,preprintnumbers,amsmath,amssymb,prb,apsrev4-2]{revtex4-2}
\usepackage{graphicx}
\usepackage{dcolumn}
\usepackage{bm}
\usepackage{SIunits}
\usepackage{verbatim}
\usepackage{placeins}
\usepackage{natbib}
\usepackage{amsmath}
\usepackage{amssymb}
\usepackage{longtable}
\usepackage{tabularx}
\usepackage{keyval}
\usepackage{multirow}
\usepackage[colorlinks,linkcolor=blue]{hyperref}

\begin{document}
\title{Multiple topological phases of magnons induced by Dzyaloshinskii-Moriya and pseudodipolar anisotropic exchange interactions in Kagome ferromagnets}
\author{Jin-Yu Ni$^{1,2}$}
\author{Xia-Ming Zheng$^{1}$}
\author{Peng-Tao Wei$^{1}$}
\author{Da-Yong Liu$^{3,1}$}
\email[]{dyliu@ntu.edu.cn}
\author{Liang-Jian Zou$^{1}$}
\email[]{zou@theory.issp.ac.cn}
\affiliation{1 Key Laboratory of Materials Physics, Institute of Solid State Physics, HFIPS, Chinese Academy of Sciences, Hefei 230031, China}
\affiliation{2 Science Island Branch of Graduate School, University of Science and Technology of China, Hefei 230026, China}
\affiliation{3 Department of Physics, School of Physical Science and Technology, Nantong University, Nantong 226019, China}
\href{http://orcid.org/0000-0003-4370-473X}{ORCiD: 0000-0003-4370-473X}

\date{\today}

\begin{abstract}
Kagome magnets naturally hosting Dirac points and flat bands exhibit novel topological phases, enabling rich interplays between interactions and topologies. The discovery of two-dimensional (2D) magnets generally coexisting with different types of magnetic interactions poses a challenge for topological magnonic manipulation. Here we investigate the topological magnon phases of 2D Kagome ferromagnet with multiple magnetic anisotropic interactions, {\it i.e.} Dzyaloshinskii-Moriya interaction (DMI) and pseudo-dipolar interaction (PDI). It is found that the different sole magnetic anisotropic interactions introduce completely distinct topological phase diagrams and topological states. The multiple topological magnon phases with high Chern number emerge due to the distinct anisotropic interactions. Moreover, the interplay of the multiple anisotropic DMI and PDI interactions involved with Dirac and flat bands controls a variety of topological phase transitions, implying greater manipulation potential. In addition, the sign reversal of thermal Hall and Nernst conductivities induced by temperature is found in particular topological phase regions, namely topological origin, relating to the energy gap and Berry curvature (Chern number) in the vicinity of magnetic phase transition from the thermal fluctuations, providing a possible explanation for the experimental puzzles.
All these results demonstrate that the novel topological magnonic properties in Kagome magnet with multiple magnetic anisotropic interactions can realize a potential platform for magnonic devices and quantum computing.
\end{abstract}

\vskip 300 pt

\maketitle

\section{Introduction}
With the discovery of topological insulators \cite{RMP82-3045,RMP83-1057}, more and more novel topological states and materials relevant to fermions have been excavated over the past decade. At the same time, topological bosons also attract much attention, such as phonon \cite{PRL95-155901,PRL100-145901,NATURE555-342}, photon \cite{PRL93-083901,PRA78-033834,RMP91-015006}, magnon \cite{PRB87-144101,JPCM28-386001,PR915-1,JAP129-151101,PRA9-024029,PRL104-066403,JPCM29-385801,PRL117-227201,PRL115-106603,PRL103-047203,JPCM34-495801}, {\it etc}. Magnon is the quantum of spin wave which is a collective excitation of a magnetic ordered system. The propagation of magnons does not depend on the movement of electrons so that it is free of Joule heat and can propagate in insulating materials \cite{NP11-453,Science366-1125}. In order to control the transport properties of magnons in a directional way, there is an urgent need to study topological magnons which just propagate along the edge regardless of the material's geometry. The magnon thermal Hall effect is first observed experimentally in an insulating ferromagnet Lu$_{2}$V$_{2}$O$_{7}$ with pyrochlore lattice by Onose \cite{SCIENCE329-297} which greatly promotes the application of topological magnon, and also identify the theoretical predictions by Matsumoto, Murakami \cite{PRL106-197202,PRB84-184406} and Zhang \cite{PRB87-144101} {\it et al}. Along with a great deal of the theoretical proposals and experimental confirmation of topological magnon phases, magnonics and topological magnonics consequently emerged \cite{JAP129-151101,PR905-1,PR915-1,JPCM33-413001,ARCMP13-171}.

Topological magnons have been proposed in different types of crystal lattices theoretically and experimentally, such as honeycomb \cite{JPCM28-386001,JPCM36-255801,PRB95-014435,PE135-114984,PRL127-217202}, triangular \cite{JPCM29-385801,PRB100-064412}, Kagome \cite{PRL115-147201,PRB90-024412,PRB104-144422,PRB101-100405,PRB97-094412,NC5-4815,PRB103-054405,SCPMA63-107511}, Lieb \cite{JPCM27-166003,JMMM417-208}, honeycomb-Kagome \cite{JPCM34-505801}, triangular-Kagome \cite{PRB107-024408} lattices, {\it etc}. 
Correspondingly, rich magnon topological phases or states associated with non-zero Berry curvature are also unveiled, including Z$_{2}$ topological magnon insulator \cite{ARCMP13-171,PR915-1}, topological magnon Chern insulator \cite{ARCMP13-171,PR915-1}, topological magnon Dirac \cite{PRB94-075401,PRX8-011010} and Weyl \cite{NC7-12691,PRL117-157204,PRB95-224403} semimetals, and high-order topological magnon insulator \cite{ARCMP13-171,PR915-1}, {\it etc}.
Among them, topological magnon insulator, analogue of topological insulator of electronic version, naturally exhibits robust non-trivial surface/edge/end states in the energy gap protected by the symmetries of the system.
Analogously, the topological Dirac/Weyl magnons, possess linear Dirac dispersion in the honeycomb and Kagome lattices.
In real magnetic materials, the corresponding topological phases are also observed experimentally, such as for the 3$d$ transition-metal compounds, the topological magnon insulator in Lu$_{2}$V$_{2}$O$_{7}$ \cite{SCIENCE329-297,PRL113-047202,PRB89-134409}, the Dirac and nodal-line magnons in Cu$_{3}$TeO$_{6}$ \cite{PRL119-247202,NC9-2591}, 
and the various topological magnon phases in Mn$_{5}$Ge$_{3}$ \cite{NC14-7321}, CoTiO$_{3}$ \cite{PRX10-011062,NC12-3936}, CrI$_{3}$ \cite{PRX8-041028}, CrSiTe$_{3}$ and CrGeTe$_{3}$ \cite{SciAdv7-eabi7532}.
Especially, the topoloigcal magnon insulator in a metal-organic framework compound Cu[1,3-benzenedicarboxylate(bdc)] ferromagnet with the Kagome lattice \cite{PRL115-147201}, and the Dirac magnon in Kagome compound TbMn$_{6}$Sn$_{6}$ \cite{NC15-1592}, are also reported.
In addition, topological magnons are extensively observed in different types of transition-metal compounds, such as 4$d$ $\alpha$-RuCl$_{3}$ \cite{PRL120-217205,PRB99-085136}, 5$d$ Na$_{2}$IrO$_{3}$ \cite{PRL108-127204} and Sr$_{3}$Ir$_{2}$O$_{7}$ \cite{npjQM4-23}, and 4$f$ elemental metal Gd \cite{PRL128-097201}, {\it etc}.

However, the microscopic origins of the magnonic band topology in these compounds are diverse due to the distinct magnetic interactions in massive magnetic materials. Generally, the symmetric interactions like the Heisenberg interaction can not directly induce topological magnons in common honeycomb and Kagome lattices, because the degeneracy of the magnonic bands can not be broken. On the contrary, the antisymmetric exchange interactions such as Dzyaloshinskii-Moriya interaction (DMI) \cite{JPCS4-241,PR120-91}, Kitaev interaction \cite{AK321-2}, dipole-dipole interaction \cite{PRL124-077201}, and pseudo-dipolar interaction (PDI) \cite{PR52-1178,PRB59-1079,PRL102-017205}, can introduce numerous topological magnon phases and novel properties. 
The antisymmetric interactions of magnons are like the spin-orbit coupling (SOC) for electrons which is the primary origin of the various anisotropic and topological properties. Similar to the DMI essentially originating from the SOC \cite{PR120-91}, the PDI is from the combined effects of the SOC and orbital quenching which widely exist in the 3$d$, 4$d$ and 5$d$ transition-metal compounds \cite{PRL102-017205,JPCM34-495801,JPCM36-255801}. Unlike the DMI, which requires inversion symmetry breaking, the PDI is derived only from the orbital physics of the transition-metal compounds with the unfilled 3$d$, 4$d$, 5$d$, and 4$f$ electrons \cite{PRL102-017205}. In comparison, both as the bond-dependent interactions, the Kitaev interaction can be mapped to the linear term of the PDI \cite{PRL125-217202}. That is, the PDI is a more ubiquitous anisotropic interaction in comparison with the Kitaev interaction.
Naturally, in a magnetic system there indeed exhibits multiple anisotropic interactions and the interplay of them still remains unclear. Whether the multiple anisotropic interactions are competitive or cooperative remains controversial. 
On the other hand, the Kagome materials have attracted much interest on its exotic electronic states, {\it e.g.} charge density wave (CDW), unconventional superconductivity, and novel band topology, owning to the coexistence of the Dirac and flat bands \cite{PRB89-134409}, as well as a joint lattice geometry with both the triangular and hexagonal arrangement of sites \cite{PRB89-134409}.  
As a consequence, the magnonic bands in Kagome magnets naturally host Dirac and flat bands, enabling complex interplays among various magnetic interactions and topology. 
Despite of some previous studies of topological magnons, the Kagome magnets generally exhibiting different types of magnetic interactions pose a challenge for understanding the underlying mechanisms of the topological magnon phases.
Motivated by the above reasons, we investigate the effect of the combined nearest-neighboring (N.N.) and next nearest-neighboring (N.N.N.) symmetric Heisenberg interactions, and the antisymmetric DMI and PDI on topological magnons in a two-dimensional (2D) Kagome ferromagnet. 
By calculating a series of features of topology and the topological phase diagrams, we find that the multiple anisotropic interactions have a strong coupling, and the interplay of them manifests either competition or cooperation, depending on the sign and strength of the interactions.
In addition, the topological phase transition and the tunable transport properties, such as sign change of the thermal Hall and Nernst conductivities, demonstrate a regulating effect, which has potential applications on magnonic devices \cite{NP11-453,JPDAP43-264001}.
These exotic properties enable multiple tunable topological states, which can substantially be achieved in real transition-metal materials.
The rest of this paper is organized as follows: first, the model and method are introduced in section-II; then the results and discussions are given in section-III; and the final section is devoted to the remarks and conclusions.

\section{Model and Method}
\begin{figure}[htbp]
\hspace*{-2mm}
\centering
\includegraphics[trim = 0mm 0mm 0mm 0mm, clip=true, angle=0, width=0.6 \columnwidth]{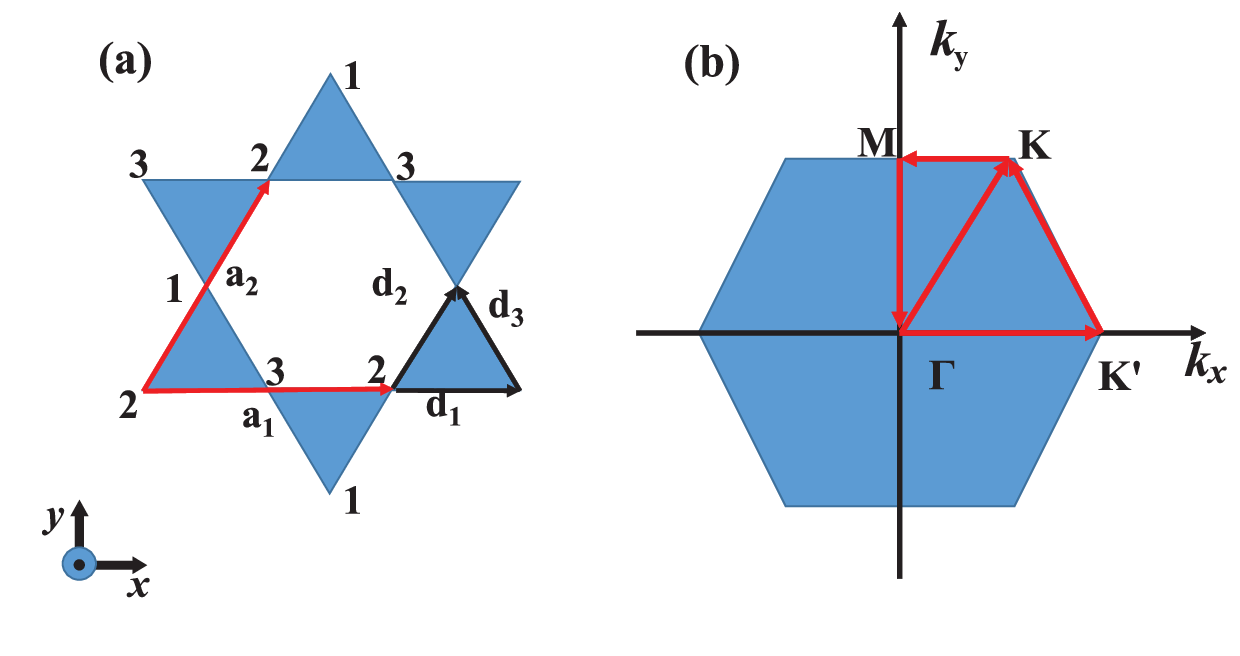}
\caption{(Color online)(a) The Kagome lattice with three inequivalent atoms with sites 1, 2 and 3, $\mathbf{a_{i}}$ is the lattice vector and $\mathbf{d_{i}}$ is the nearest-neighboring vector. (b) The corresponding Brillouin zone (BZ), and its high-symmetry $\mathbf{k}$-points and $\mathbf{k}$-paths.}
\label{Fig1}
\end{figure}

In Kagome lattice, there are three different atoms at the corners of triangle, we labeled them 1, 2, 3 as shown in Fig.~\ref{Fig1}, $\mathbf{a_{1,2}}$ are the lattice vectors and $\mathbf{d_{1,2,3}}$ are the vectors between the N.N. atoms. 
Obviously, it exhibits both the triangular and hexagonal types of lattice geometry, implying intrinsic frustration and competitive physics. 
%
Here for a two dimensional (2D) ferromagnet on Kagome lattice, the Hamiltonian of the magnetic interactions can be written as \cite{JPCM34-495801,JPCM36-255801}
\begin{equation}\label{Eq1}
    \begin{split}
  H=&-J_{1}\sum\limits_{\langle i,j\rangle}\mathbf{S_{i}}\cdot \mathbf{S_{j}}-J_{2}\sum\limits_{\langle\langle i,j\rangle\rangle}\mathbf{S_{i}}\cdot \mathbf{S_{j}}\\\\
  &+\sum\limits_{\langle i,j\rangle}\mathbf{D_{ij}}\cdot (\mathbf{S_{i}}\times \mathbf{S_{j}})-\sum\limits_{\langle i,j\rangle}F_{ij}(\mathbf{S_{i}}\cdot \mathbf{e_{ij}})(\mathbf{S_{j}}\cdot \mathbf{e_{ij}})\\\\
  &-K_{s}\sum\limits_{i}\mathbf{S_{i}^{z}}\cdot \mathbf{S_{i}^{z}}-g\mu_{B}B\sum\limits_{i}\mathbf{S_{i}^{z}},
    \end{split}
\end{equation}
where $\langle i,j\rangle$ denotes the N.N. sites, $\langle\langle i,j\rangle\rangle$ denotes the N.N.N. sites. The first two terms are the N.N. and N.N.N. Heisenberg exchange interactions with the corresponding exchange constants $J_{1}$ and $J_{2}$. The third term is the N.N. DMI which arises from the lack of inversion symmetry. Here we just take the out-of-plane direction ($\textbf{\textit{z}}$) component into account, $\mathbf{D_{ij}}=\nu_{ij}D_{z}\mathbf{\hat{z}}$, where $\nu_{ij}=\pm 1$ correspond to the anticlockwise and clockwise directions of the N.N. couplings in a triangle formed by sites 1, 2 and 3 as displayed in Fig.~\ref{Fig1}(a), and $\mathbf{\hat{z}}$ is the unit vector along the $\textbf{\textit{z}}$ direction. The fourth term $F_{ij}$ is the PDI \cite{JPCM34-495801,JPCM36-255801}, here we only consider the N.N. PDI ($F$), and $\mathbf{e_{ij}}$ is the unit vector connecting sites $i$ and $j$ chosen from three N.N. vectors $\mathbf{d_{1,2,3}}$. The last two terms, {\it i.e.} the single ion anisotropic (SIA) $K_{s}$ and magnetic field $\mathbf{B}$ terms, are both oriented to the out-of-plane direction.

We perform the standard linear Holstein-Primakoff (HP) transformation \cite{PR58-1098} on the Hamiltonian $H$,
\begin{equation}\label{Eq2}
S_{i}^z=S-a_{i}^{\dagger} a_{i},S_{i}^{+}=\sqrt{2S}a_{i},S_{i}^{-}=\sqrt{2S}a_{i}^{\dagger},
\end{equation}
where $a_{i}^{\dagger}$ and $a_{i}$ are the bosonic creation and annihilation operators, and $S_{i}^{\pm}$=$S_{i}^{x}$$\pm$$S_{i}^{y}$ denote the spin raising and lowering operators. After the Fourier transformation into momentum space, we get a $6 \times 6$ Hamiltonian $H=\frac{1}{2}\sum_{\mathbf{k}}\Psi_{\mathbf{k}}^{\dagger}H(\mathbf{k})\Psi_{\mathbf{k}}$ with $\Psi_{\mathbf{k}}^{\dagger}$=$(a_{1, \mathbf{k}}^{\dagger}, a_{1, -\mathbf{k}}, a_{2, \mathbf{k}}^{\dagger}, a_{2, -\mathbf{k}}, a_{3, \mathbf{k}}^{\dagger}, a_{3, -\mathbf{k}})$. Then the Hamiltonian $H(\mathbf{k})$ can be described by
\begin{equation}\label{Eq3}
H(\mathbf{k})=\left(
    \begin{array}{cccccc}
    h_{0} & 0 & f_{1}(\mathbf{k})  & g_{1+}(\mathbf{k}) & f_{2}(\mathbf{k}) & g_{2+}(\mathbf{k})\\
    0 & h_{0} & g_{1-}(\mathbf{k}) & f_{1}^{\ast}(\mathbf{k}) & g_{2-}(\mathbf{k}) & f_{2}^{\ast}(\mathbf{k})\\
    f_{1}^{\ast}(\mathbf{k}) &  g^{\ast}_{1-}(\mathbf{k}) & h_{0} & 0 & f_{3}(\mathbf{k}) & g_{3+}(\mathbf{k})\\
    g^{\ast}_{1+}(\mathbf{k}) & f_{1}(\mathbf{k}) & 0 & h_{0} & g_{3-}(\mathbf{k}) & f_{3}^{\ast}(\mathbf{k}) \\
    f_{2}^{\ast}(\mathbf{k}) &  g^{\ast}_{2-}(\mathbf{k}) & f_{3}^{\ast}(\mathbf{k}) & g^{\ast}_{3-}(\mathbf{k}) & h_{0} & 0\\
    g^{\ast}_{2+}(\mathbf{k}) & f_{2}(\mathbf{k}) & g^{\ast}_{3+}(\mathbf{k}) & f_{3}(\mathbf{k}) & 0 & h_{0}
    \end{array}\right),
\end{equation}
where $h_0=4(J_{1}+J_{2}+K_{s}/2)S+g\mu_{B}B+E_{0}$ and $E_{0}$ is ground state energy, \\
$f_{1}(\mathbf{k})=-(J_{1}+\frac{1}{2}F-iD_{z})S\times 2cos(\mathbf{k}\cdot \mathbf{d_{2}})-J_{2}S\times2cos(\mathbf{k}\cdot (\mathbf{d_{3}}-\mathbf{d_{1}}))$, \\
$f_{2}(\mathbf{k})=-(J_{1}+\frac{1}{2}F+iD_{z})S\times 2cos(\mathbf{k}\cdot \mathbf{d_{3}})-J_{2}S\times 2cos(\mathbf{k}\cdot (\mathbf{d_{1}}+\mathbf{d_{2}}))$, \\
$f_{3}(\mathbf{k})=-(J_{1}+\frac{1}{2}F-iD_{z})S\times 2cos(\mathbf{k}\cdot \mathbf{d_{1}})-J_{2}S\times 2cos(\mathbf{k}\cdot (\mathbf{d_{2}}+\mathbf{d_{3}}))$, \\
$g_{1\pm}(\mathbf{k})=\frac{1}{2}FS(e^{i(\pm 2\theta_{2}+\mathbf{k}\cdot \mathbf{d_{2}})}+e^{i(\pm 2(\theta_{2}+\pi)-\mathbf{k}\cdot \mathbf{d_{2}})})$, \\
$g_{2\pm}(\mathbf{k})=\frac{1}{2}FS(e^{i(\pm 2\theta_{3}+\mathbf{k}\cdot \mathbf{d_{3}})}+e^{i(\pm 2(\theta_{3}+\pi)-\mathbf{k}\cdot \mathbf{d_{3}})})$, \\
$g_{3\pm}(\mathbf{k})=\frac{1}{2}FS(e^{i(\pm 2\theta_{1}+\mathbf{k}\cdot \mathbf{d_{1}})}+e^{i(\pm 2(\theta_{1}+\pi)-\mathbf{k}\cdot \mathbf{d_{1}})})$, \\
and $\theta_{i}$ is the angle between $\mathbf{d_{i}}$ and $\textbf{\textit{x}}$ direction, and $\mathbf{d_{1}}=(\frac{1}{2},0),\mathbf{d_{2}}=(\frac{1}{4},\frac{\sqrt{3}}{4}),\mathbf{d_{3}}=\mathbf{d_{2}}-\mathbf{d_{1}}$.
Then we can diagonalize the magnon Hamiltonian $H(\mathbf{k})$, which satisfies the generalized eigenvalue problem, and obtain the energy spectrum $E(\mathbf{k})$ of the magnon.

Furthermore, we use the Chern number to characterize the non-trivial topological magnons in topological magnon insulator, due to the gap opening induced by the anisotropic exchange interactions DMI ($D_{z}$) and PDI ($F$) in Kagome lattice. 
Chern number as one of the topological invariants, is an integration of Berry's curvature on the whole Brillouin zone (BZ) \cite{PRSLA392-45}. 
The Chern number of one band ($C_{n}$ for the $n$-th band) can be defined as \cite{PRB87-174427}
\begin{equation}\label{3}
C_{n}=\frac{1}{2\pi}\int_{BZ}d^{2}kB_{n}(\mathbf{k})
\end{equation}
$B_n(\mathbf{k})$ is the Berry curvature of the $n$-th magnon band in momentum space, 
\begin{equation}\label{4}
B_n(\mathbf{k})=i\epsilon_{\mu \nu} \operatorname{Tr}\left[(1-Q_n)\left(\partial_{k_\mu} Q_n\right)\left(\partial_{k_\nu} Q_n\right)\right],
\end{equation}
which is different from the fermionic system due to the bosonic diagonalization. $Q_{n}$ is a projection operator in the vector space $Q_n=T_{\mathbf{k}} \Gamma_n \sigma_3 T_{\mathbf{k}}^{\dagger} \sigma_3$. $T_{\mathbf{k}}$ is a paraunitary matrix satisfied $T_{\mathbf{k}}^{\dagger} H_{\mathbf{k}} T_{\mathbf{k}}=\left[\begin{array}{ll}
E_{\mathbf{k}} & \\
& E_{-\mathbf{k}}
\end{array}\right]$, $\sigma_3$ is the Pauli matrix, and $\Gamma_n$ is a diagonal matrix with $+$1 for the $n$-th diagonal component and zero otherwise \cite{PRB87-174427}. In addition, an efficient method \cite{JPSJ74-1674} for calculating Berry curvature and Chern number is also adopted, and the results are cross-checked to verify the reliability and accuracy of the numerical calculations.

\section{Results and Discussions}

\subsection{Classical Ground State}
\begin{figure}[htbp]
\hspace*{-2mm}
\centering\includegraphics[trim = 0mm 0mm 0mm 0mm, clip=true, angle=0, width=0.6 \columnwidth]{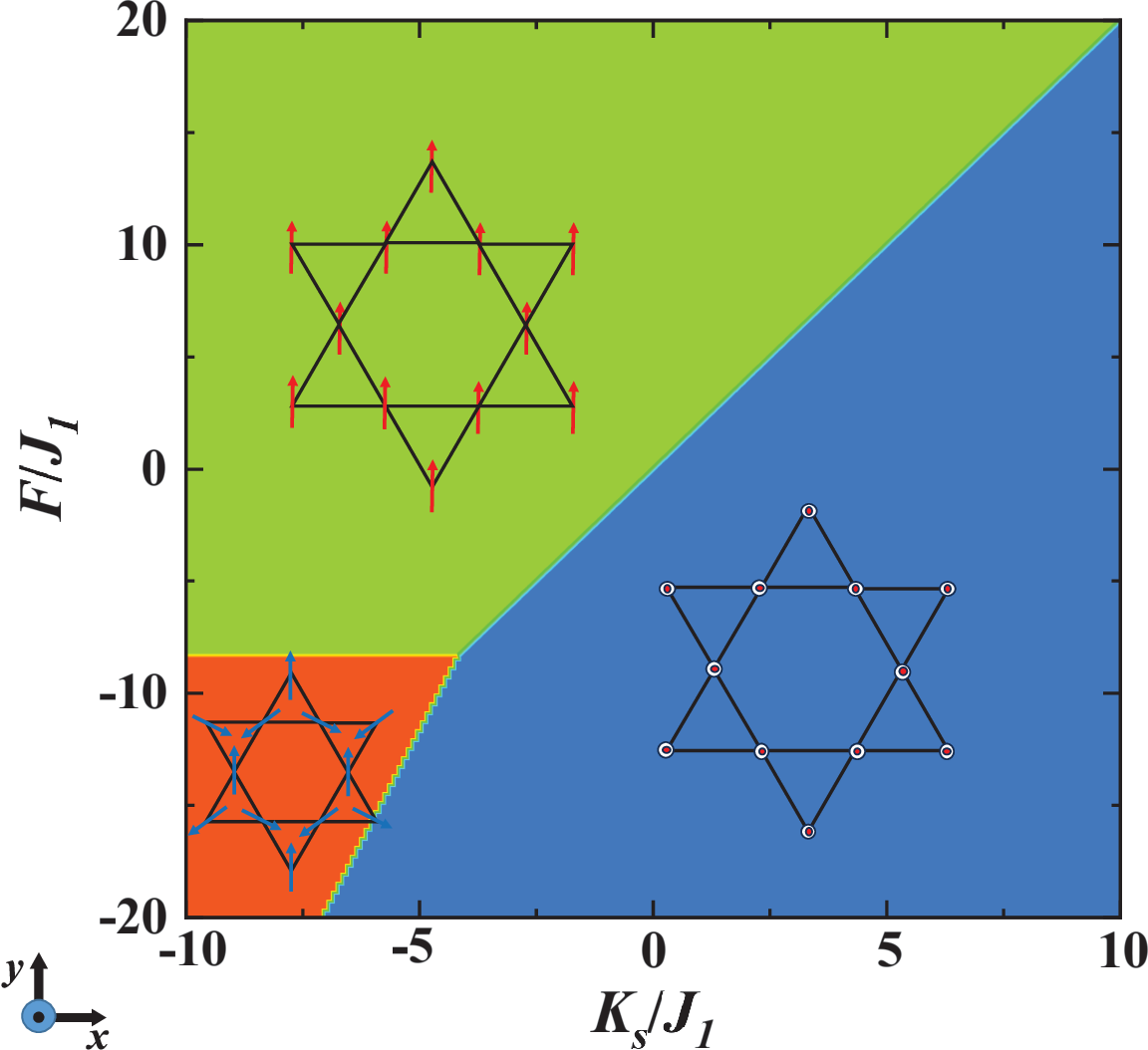}
\caption{(Color online) $K_{s}$-$F$ phase diagram of classical ground state with fixed parameters $B=0$, $J_{1}$=1, $J_{2}$=0.5 and $D_{z}$=0.2. Insets are the corresponding stable spin configurations.}
\label{Fig2}
\end{figure}

In order to obtain the magnon spectrum, the equilibrium state of the spin configurations of the Kagome lattice should be determined. 
Here we focus on the classical spin ground state, which is valid for large $S$ value.
Considering the complex magnetic interactions in Eq.~\ref{Eq1}, for simplicity, we calculated the magnetic ground state energies of several conventional spin configurations with the Heisenberg exchange constants $J_{1}$ and $J_{2}$ $>$ 0 (mainly a ferromagnetic (FM) solution).
However, the anisotropic exchange interactions $D_{ij}$ and $F_{ij}$ may alter the FM configuration to the antiferromagnetic (AFM) one due to the competing interactions. 
For convenience, we fix several interactions, {\it e.g.} the following parameters $J_{1}$=1, $J_{2}$=0.5 and $D_{z}$=0.2, to investigate the $K_{s}$-$F$ phase diagram for the classical spins. Then we obtain three stable spin configurations, as shown in Fig.~\ref{Fig2}. 
In the magnetic phase diagram, the blue region indicates a FM order with spins aligning along the $\textbf{\textit{z}}$ direction, {\it i.e.} perpendicular or out-of-plane FM, and the total energy $E_{FM-z}=-6(J_{1}+J_{2}+K_{s})NS^{2}$, where $N$ is the number of sites per unit cell. The green region is a FM state in which spins lie in the $x$-$y$ plane (in-plane FM) with $E_{FM-xy}=-6(J_{1}+J_{2}+F/2)NS^{2}$. And the red one is a 120$^{\circ}$ noncollinear spin configuration in the $x$-$y$ plane (in-plane 120$^{\circ}$ AFM) with $E_{AFM}=-3(-J_{1}-J_{2}+\sqrt{3}D_{z}+F/2)NS^{2}$.

Here we need to emphasize that, the strength of the anisotropic exchange interactions (DMI and PDI) and SIA in the realistic magnetic materials should be taken into account. Evidently, both the theoretical estimations and experimental observations suggest an extensive existence of the strong SIA, DMI and PDI, and the details can be found in the previous studies \cite{JPCM34-495801,JPCM36-255801}. Particularly, a giant ratio between anisotropic interactions and Heisenberg interactions ({\it e.g.} $K_{s}/J_{1}$, $D_{z}/J_{1}$ and $F/J_{1}$) can be achieved as reported in a series of experiments of transition-metal compounds, indicating a strong magnetic anisotropy phenomenon due to the complicated charge-spin-lattice-orbital couplings in low-dimensional magnets \cite{JPCM34-495801,JPCM36-255801}. The 2D magnetic materials inherently exhibit relatively strong magnetic anisotropy to maintain stability, making them natural platforms for magnetic anisotropy. The strong SIA systems can be realistically achieved on Kagome Ising ferromagnets \cite{npjCM6-158,PRB106-115139,arXiv2412.02010} with dominant $K_{s}$ over $J_{1}$. While the PDI originating from the interplay of strong orbital ordering and SOC in the 3$d$, 4$d$, 5$d$, 4$f$ and 5$f$ transition metal compounds, corresponds to the so-called Kagome Kitaev magnets \cite{JPCM29-493002,PRB89-014414,PRB98-134437}, exhibiting strong magnetic exchange anisotropy $F$ even dozens of times stronger than $J_{1}$ \cite{JPCM36-255801}.

Notice that we only focus on the magnons of the Kagome ferromagnet. In order to ensure that the ground state always maintains the FM-$z$ configuration, corresponding to the blue region in Fig.~\ref{Fig2}, unless otherwise stated, we fixed the parameters $K_{s}=10$ and $B=0$ in the rest of this paper. 
Indeed, it can be tuned by the control of the combined SIA and external magnetic field in the realistic materials. Then we can investigate the magnon spectrum and its involved topological properties based on the spin ground state of the Kagome ferromagnet.

\subsection{Magnonic Band Structure}
\begin{figure}[htbp]
\hspace*{-2mm}
\centering\includegraphics[trim = 0mm 0mm 0mm 0mm, clip=true, angle=0, width=0.8 \columnwidth]{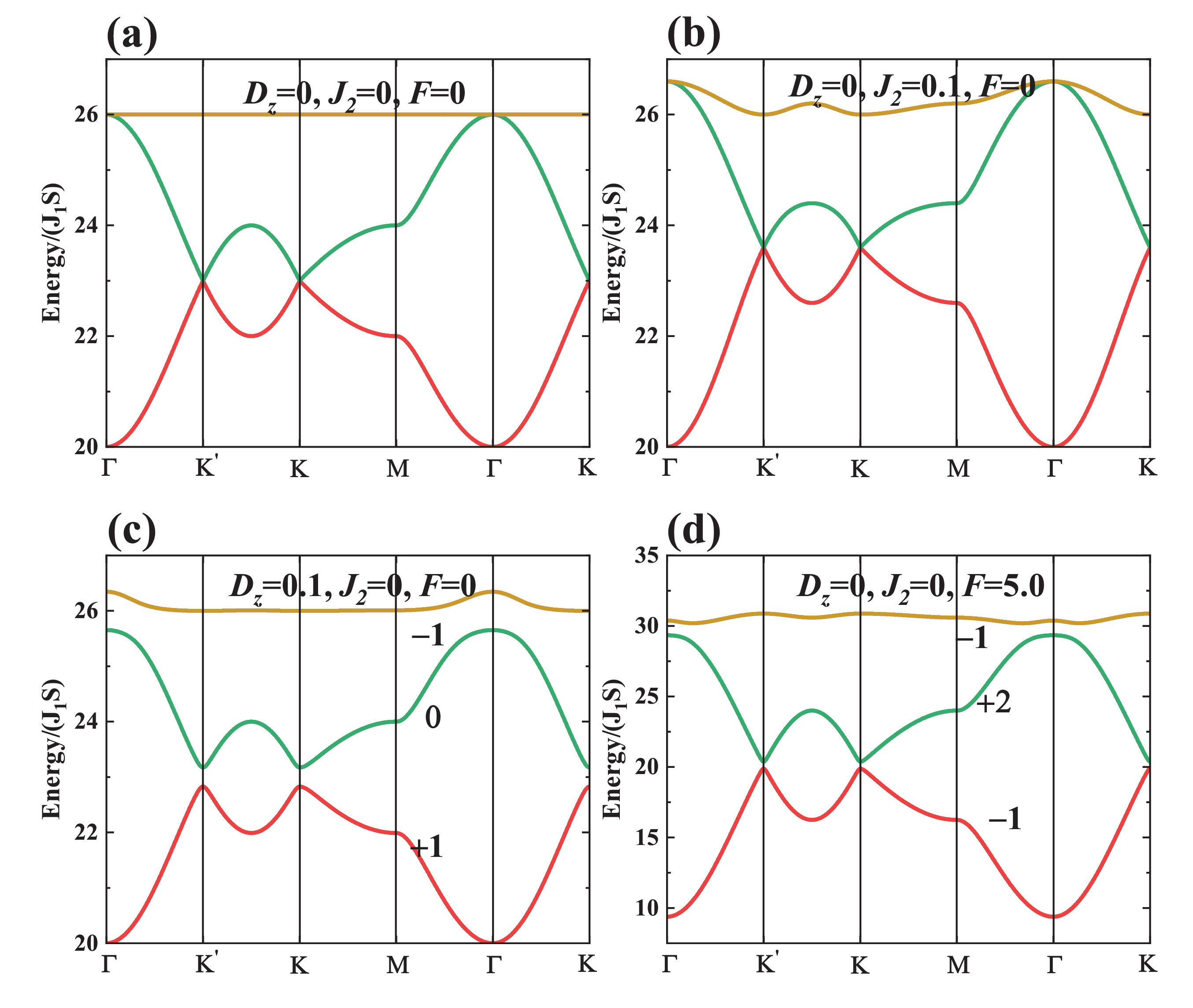}
\caption{(Color online) Band structures of magnons with different parameters (a) $J_{2}$=$D_{z}$=$F$=0, (b) $J_{2}$=0.1 and $D_{z}$=$F$=0, (c) $J_{2}$=$F$=0 and $D_{z}$=0.1, (d)$J_{2}$=$D_{z}$=0 and $F$=5. The Chern numbers $C_{n}$ of each band from top to bottom are marked.}
\label{Fig3}
\end{figure}

Due to the presence of the multiple interactions in spin Hamiltonian Eq.~\ref{Eq1}, the first is to distinguish the contributions of individual magnetic interaction terms.
Subsequently, the influence of the different magnetic exchange interactions on the magnon spectra are investigated for comparison, as demonstrated in Fig.~\ref{Fig3}(a)-(c).
Figure.~\ref{Fig3} (a) shows the band dispersion of magnons for the N.N. Heisenberg model with only $J_{1}$ on the Kagome lattice. It is found that there is a perfect flat band on top (upper band) which relates to a localized magnetic excitation. And the other two bands on bottom (middle and lower bands) have a linear Dirac dispersion at $\mathbf{K}$ and $\mathbf{K}^{\prime}$ points. 
The resulting magnonic band structure is consistent with the electronic version, both originating from the geometric features of the Kagome lattice.
The N.N.N. exchange interaction $J_{2}$ can enhance the dispersion on top but can not open a gap at the linear crossing, and the system is still topologically trivial in Fig.~\ref{Fig3}(b). 
The DMI $D_{z}$ and PDI $F$ all can induce topological magnons individually with band gap opening, as shown in Fig.~\ref{Fig3}(c) and (d).
Notice that ($-$1, 0, $+$1) and ($-$1, $+$2, $-$1) are the sets of Chern numbers of the upper, middle and lower bands corresponding to the DMI and PDI cases, respectively.
Then we will discuss the topological magnons related to the N.N.N. exchange interaction ($J_{2}$), DMI ($D_{z}$), and PDI ($F$) in the following.

\subsection{Topological Phase Diagram with DMI}
We first investigate the influence of the anisotropic DMI on the magnon spectrum.
Considering that the non-trivial topological states are mainly induced by the gap opening at Dirac points, the topological phase transition can be determined by the critical condition of the band gap closing. Therefore, we focus on the high symmetry point $\mathbf{K}$=($\frac{\pi}{3}$, $\frac{\sqrt{3}\pi}{3}$) or $\mathbf{K'}$=($\frac{2\pi}{3}$, 0) where Dirac point occurs.
\begin{figure}[htbp]
\hspace*{-2mm}
\centering\includegraphics[trim = 0mm 0mm 0mm 0mm, clip=true, angle=0, width=0.8 \columnwidth]{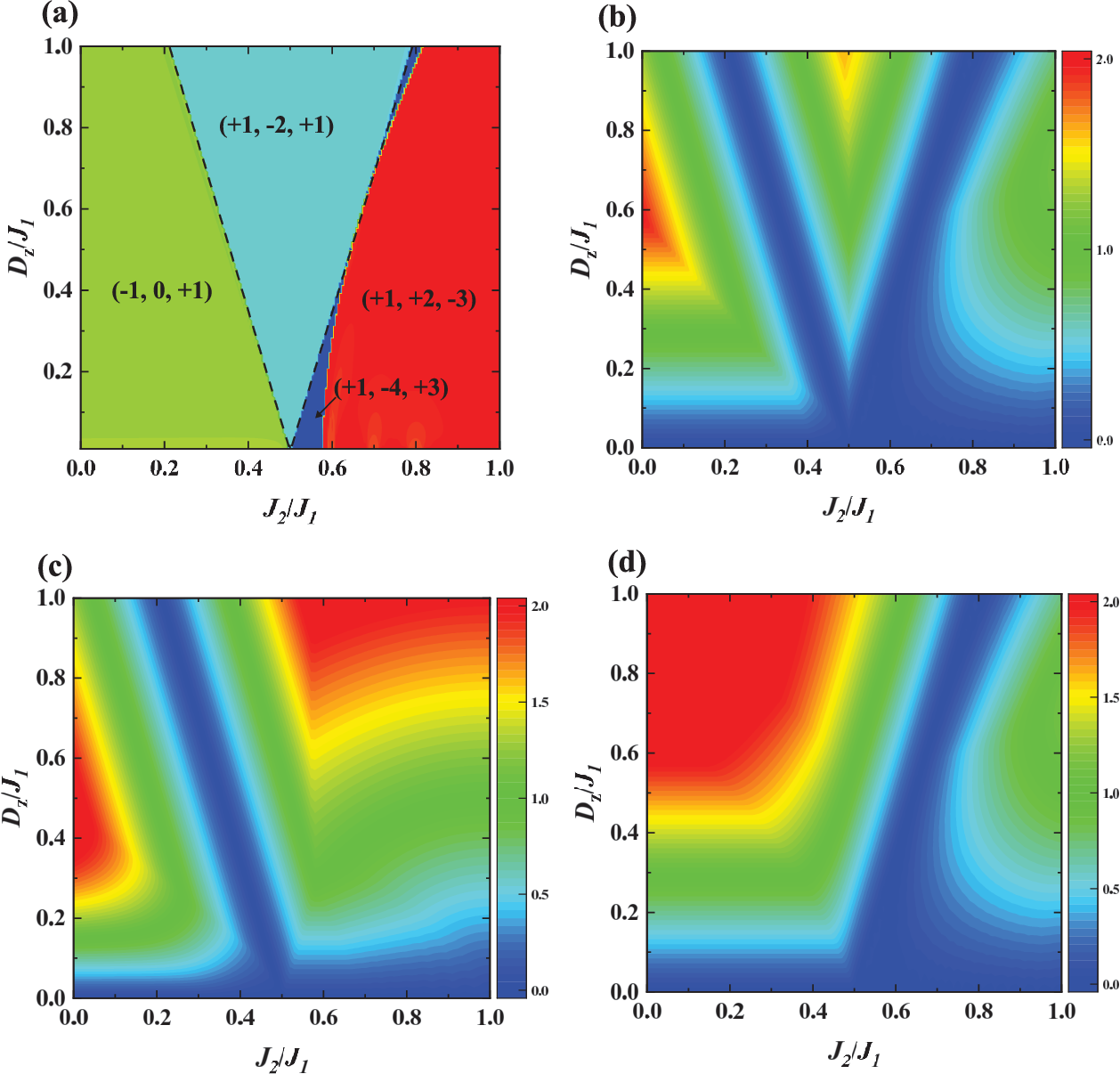}
\caption{(Color online) (a) $J_{2}$-$D_{z}$ topological phase diagram. The Chern numbers of the upper (1-st), middle (2-nd) and lower (3-rd) bands are marked in color regions to denote different topological phases. (b) The corresponding minimum band gap $\Delta_{Min}$ between the different magnon bands, {\it i.e.} Min($\Delta_{1,2}$, $\Delta_{2,3}$). (c) and (d) are the band gaps between the different magnon bands, $\Delta_{1,2}$ and $\Delta_{2,3}$, for the corresponding topological phase diagram, respectively. Note that the parameter $F=0$ in the absence of PDI.}
\label{Fig4}
\end{figure}
In the absence of PDI, the Hamiltonian $H(\mathbf{k})$ with a 6$\times$6 matrix in Eq.~\ref{Eq3} can be simplified to a 3$\times$3 form. We can get a 3$\times$3 Hamiltonian with $\Psi_{\mathbf{k}}^{\dagger}$ $=$ ($a_{1, \mathbf{k}}^{\dagger}$, $a_{2, \mathbf{k}}^{\dagger}$, $a_{3, \mathbf{k}}^{\dagger}$) as follows
\begin{equation}
H=\left(
    \begin{array}{ccc}
    A & C^{\ast} & -C\\
    C & A & -C^{\ast}\\
    -C^{\ast} & -C & A
    \end{array}\right)
\end{equation}
where $A=4(J_{1}+J_{2}+K_{s}/2)S+g\mu_{B}B+E_{0}$, $C=(J_{1}-2J_{2}+iD_{z})S$. Then three eigenvalues $\varepsilon_{1,2,3}$ can be obtained by diagonalizing the $H$ matrix. Thus the critical condition of topological phase transition is determined by $D_{z}=\sqrt{3}|J_{1}-2J_{2}|$ for $S$=1, as denoted by the dashed lines in Fig.~\ref{Fig4}(a), consistent with the previous study \cite{PRB90-024412}. 
As a consequence, there are mainly three topological phases as marked by ($-$1, 0, $+$1), ($+$1, $-$2, $+$1) and ($+$1, $+$2, $-$3).
Notice that the middle Chern number of the three-band Kagome system is used to label the colors of different topological phases in the topological phase diagram.
It should be emphasized that since the chirality of the DMI is not broken, the phase diagram corresponding to the negative $D_{z}$ is symmetrical to that of positive $D_{z}$, which is not plotted in the topological phase diagram of Fig.~\ref{Fig4}(a).

\begin{figure}[htbp]
\hspace*{-2mm}
\centering\includegraphics[trim = 0mm 0mm 0mm 0mm, clip=true, angle=0, width=0.8 \columnwidth]{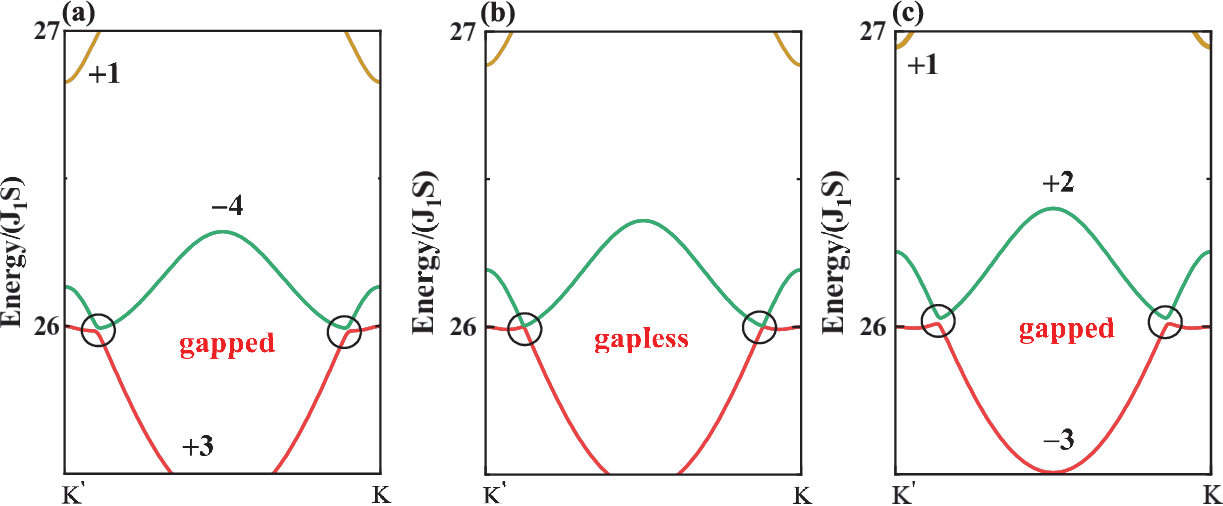}
\caption{(Color online) Enlarged band structures with fixed $D_{z}$=0.2, and $J_{2}$=0.58 (a), 0.59 (b) and 0.6 (c), respectively. The black circles depict the gap closing and opening. The Chern numbers of the corresponding upper, middle and lower bands are marked.}
\label{Fig5}
\end{figure}
However, not all the band inversions involved with the topological phase transition occur at $\mathbf{K}$ or $\mathbf{K'}$ point, but may emerge far away from the high symmetry points. In order to check the details of the topological phase transition, we calculated the band gaps between different bands over the whole Brillouin zone (BZ) as shown in Fig.~\ref{Fig4}(c) and (d). Due to the three-band character of the Kagome lattice, there are two kinds of band gaps, {\it i.e.} $\Delta_{1,2}$ between the upper and middle bands, and $\Delta_{2,3}$ between the middle and lower bands.
It is found that the topological phase diagram corresponds to the minimum band gap between the different magnon bands, {\it i.e.} Min($\Delta_{1,2}$, $\Delta_{2,3}$), as displayed in Fig.~\ref{Fig4}(b).
Note that a refined $k$-mesh can provide band gaps less than 0.001$J_{1}$.
This can be easily understood that the topological phase transition always goes with band closing and band reopening.

Especially, there really exit a new topological phase (blue region in phase diagram of Fig.~\ref{Fig4}(a)) with the sets of Chern numbers ($+$1, $-$4, $+$3) between $J_{2}$=0.5 and 0.6, which is away from the high symmetry $\mathbf{K}$ and $\mathbf{K'}$ points, as shown in the enlarged band structures of Fig.~\ref{Fig5}. 
It can be clearly seen that the band gaps go through a process of closing to reopening.
These results indicate that there is complex competition among these three bands.

\subsection{Topological Phase Diagram with PDI}
\begin{figure}[htbp]
\hspace*{-2mm}
\centering\includegraphics[trim = 0mm 0mm 0mm 0mm, clip=true, angle=0, width=0.8 \columnwidth]{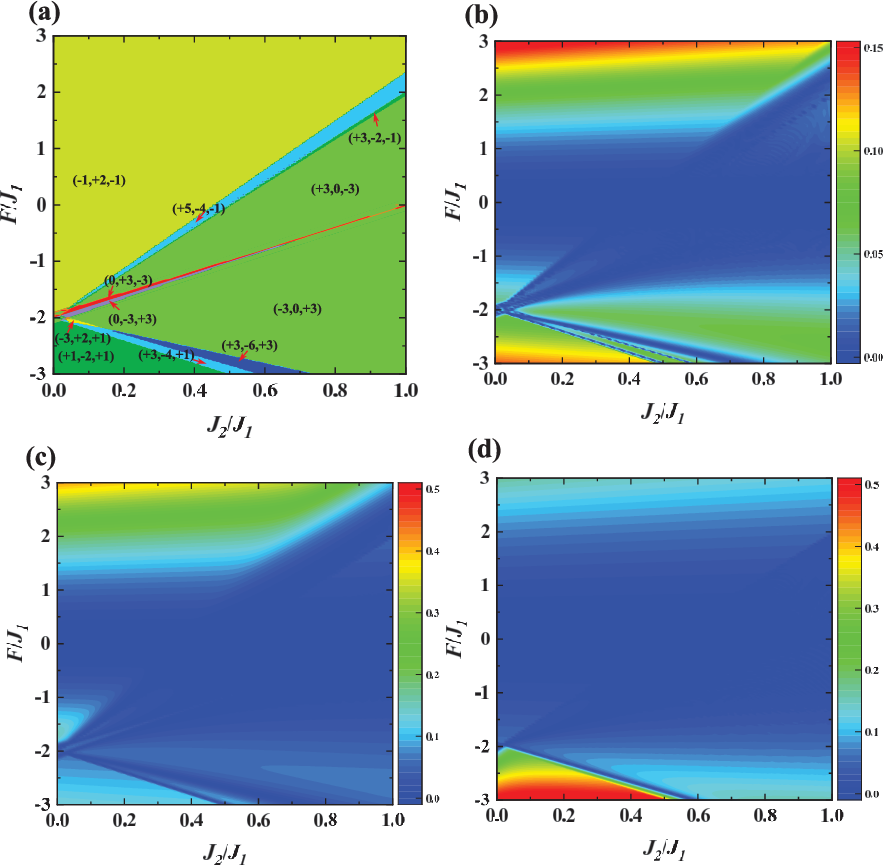}
\caption{(Color online) (a) $J_{2}$-$F$ topological phase diagram. The topological phases denoted by the Chern numbers of the upper, middle and lower bands are listed in different color regions. (b) The corresponding minimum band gap $\Delta_{Min}$ between the different magnon bands, {\it i.e.} Min($\Delta_{1,2}$, $\Delta_{2,3}$). (c) and (d) are the band gaps between the different magnon bands, $\Delta_{1,2}$ and $\Delta_{2,3}$, for the corresponding topological phase diagram, respectively. Note that the parameter $D_{z}=0$ in the absence of DMI.}
\label{Fig6}
\end{figure}

\begin{figure}[htbp]
\hspace*{-2mm}
\centering\includegraphics[trim = 0mm 0mm 0mm 0mm, clip=true, angle=0, width=1.0 \columnwidth]{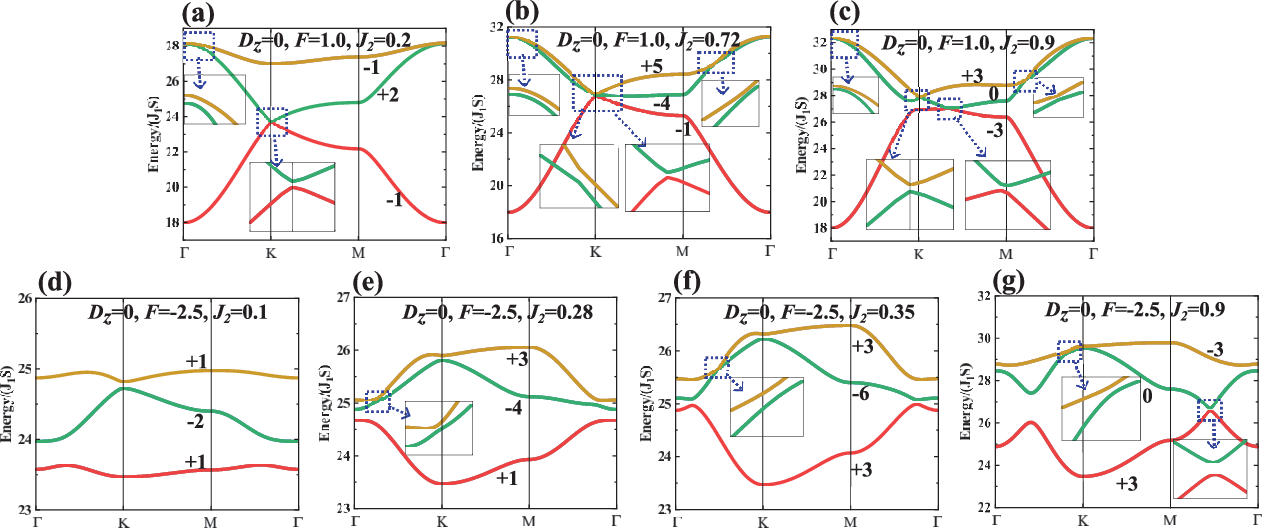}
\caption{(Color online) Band structures dependent on the different N.N.N. Heisenberg exchange interaction $J_{2}$ for the fixed PDI parameter $F$ in the absence of DMI ($D_{z}$=0), {\it i.e.} $J_{2}$=0.2 (a), $J_{2}$=0.72 (b), $J_{2}$=0.9 (c) for $F$=1.0, and $J_{2}$=0.1 (d), $J_{2}$=0.28 (e), $J_{2}$=0.35 (f), $J_{2}$=0.9 (g) for $F$=$-$2.5, respectively. The Chern number of each band is marked. Insets show the expanded view indicating the band gap opening, which are marked by the dashed boxes. 
} 
\label{Fig7}
\end{figure}

Now we turn to the PDI influence on the magnon band spectrum in Kagome lattice.
Although there has been a series of work on the topological magnonic states induced by the DMI in Kagome lattice \cite{PRB89-134409,PRL104-066403,PRB103-054405}, the possible role of the PDI is still unclear.
Here we examine the possible topological states and identify the topological phase diagram dependent on the PDI.
In comparison with the topological phases induced by the DMI, the PDI introduces a complex phase diagram with multiple topological states, as shown in Fig.~\ref{Fig6}(a).
Note that the phases with the maximal band gap $\triangle_{Max}$ $<$ 0.05$J_{1}S$ are ignored in the phase diagram. 
Therefore, the more accurate phase diagram may be more complex. However, considering that the topological phases with too small band gaps generally have no practical application because of thermal fluctuations, it is reasonable to ignore these phases.
It is noted that the partial color variations within a single phase in topological phase diagram are the minor deviations of the Chern number due to the computational accuracy in the regions with tiny band gaps originating from the competing magnetic interactions, which can be eliminated by greatly increasing the k-points mesh density. Consequently, such deviation of the Chern number in the topological phase diagram does not change our conclusion.
The corresponding band gaps between the different magnon bands, $\Delta_{1,2}$, $\Delta_{2,3}$, and their minimum value $\triangle_{Min}$=Min($\Delta_{1,2}$, $\Delta_{2,3}$), are plotted in Fig.~\ref{Fig6}(c), (d) and (b), respectively. 
It is clearly demonstrated that the topological phase transitions are essentially involved with the band gap opening and closing.
Compared with the DMI case, the $J_{2}$-$F$ phase diagram
also has ($+$1, $-$2, $+$1) topological state but no ($-$1, 0, $+$1), ($+$1, $+$2, $-$3) and ($+$1, $-$4, $+$3) ones. 
Moreover, it possesses six new topological states, {\it i.e.} ($-$3, 0, $+$3), (0, $-$3, $+$3), ($+$3, $-$6, $+$3), ($+$3, $-$2, $-$1), ($+$3, $-$4, $+$1), and ($+$5, $-$4, $-$1).
Notice that the Kitaev interaction in Kagome magnet, which can be regarded as the linear term of the PDI, had been investigated and manifested a topological phase diagram with only a few topological states \cite{CPL40-027502}.
More importantly, the positive ($F$$>$0) and negative ($F$$<$0) PDIs present completely different topological phase diagrams, that is an asymmetric one, which is different from that of the DMI case.
However, only two topological phases ($-$1, 0, $+$1) and ($+$1, 0, $-$1) are found for the coexistence of the DMI and Kitaev interaction in Kagome FM as stated in the previous study \cite{CPL40-027502}.
These results indicate that the anisotropic PDI has distinctly different effects on the topological properties than the DMI and even Kitaev interaction.
Figure~\ref{Fig7} shows the evolution of the magnonic band structures on magnetic exchange interactions in different topological phases. From the enlarged view of band opening as depicted in insets of Fig.~\ref{Fig7}(a)-(g), it is found that the three bands are completely separated and isolated, with finite energy gaps. 
The positions of the minimum band gap is located not only at conventional high symmetry points $\mathbf{K}$ and $\mathbf{K'}$, but also at high symmetry point $\Gamma$ and on the path. This implies the PDI behaves as a complex interaction rather than a common Kitaev-type one, resulting in diverse band inversions.

\subsection{Topological Phase Diagram with Multiple Anisotropic Interactions}
In addition to the sole anisotropic magnetic exchange interaction (DMI or PDI), the multiple anisotropic exchange interactions on topological magnons are also worth exploring.
As a matter of fact, the multiple interactions can be naturally realized in realistic magnetic materials with 4-5$d$/4$f$ electrons, owing to the complicated couplings of different charge-spin-lattice-orbital degrees of freedom. 
\begin{figure}[htbp]
\hspace*{-2mm}
\centering\includegraphics[trim = 0mm 0mm 0mm 0mm, clip=true, angle=0, width=0.8 \columnwidth]{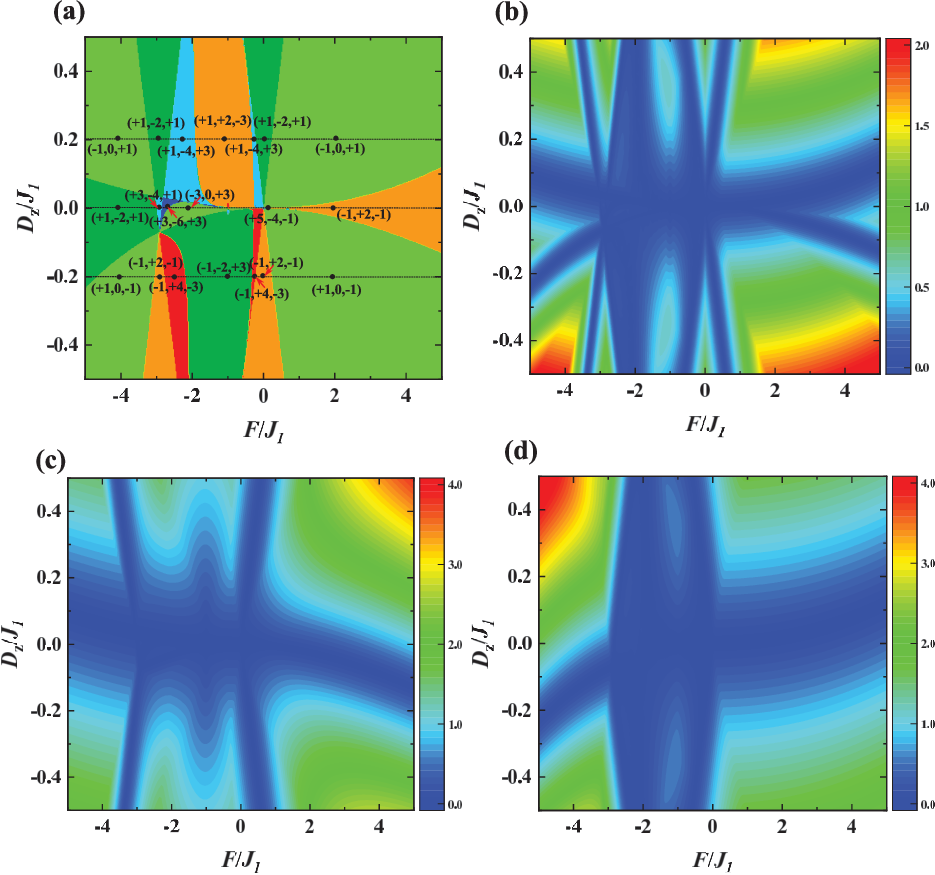}
\caption{(Color online) (a) $F$-$D_{z}$ topological phase diagram with the fixed N.N.N. exchange constant $J_{2}$=0.5. The Chern numbers of the upper, middle and lower bands in different color regions are used to mark different topological phases. (b) The corresponding minimum band gap $\Delta_{Min}$ between the different magnon bands, {\it i.e.} Min($\Delta_{1,2}$, $\Delta_{2,3}$). (c) and (d) are the band gaps between the different magnon bands, $\Delta_{1,2}$ and $\Delta_{2,3}$, for the corresponding topological phase diagram, respectively.}
\label{Fig8}
\end{figure}
In Fig.~\ref{Fig8}(a), the $F$-$D_{z}$ topological phase diagram with the fixed N.N.N. exchange parameter $J_{2}$=0.5 is shown with multiple and different topological Chern numbers. 
The interplay between the DMI and PDI indeed induces the intense band inversions accompanied by the band gap closing and reopening, leading to multiple non-trivial topological states \cite{JPCM34-495801,JPCM36-255801}.
The band gaps between the different magnon bands, $\Delta_{1,2}$ and $\Delta_{2,3}$, are also displayed in Fig.~\ref{Fig8}(c) and (d), respectively.
These three bands are coupled together with complex band gaps due to multiple anisotropic interactions, indicating the existence of competition and cooperation among these magnetic interactions.
Meanwhile rich topological phase transitions emerge at the boundaries of different topological phases, corresponding to the distribution of the minimum band gap $\triangle_{Min}$, as plotted in Fig.~\ref{Fig8}(b). 
In fact, this also means that in the case of multiple anisotropic interactions, the system can achieve rich topological regulations.
For example, the thermal Hall conductivity and magnon Nernst conductivity often significantly change in the phase transition points.
Obviously, the original symmetric topological phase diagram of the positive ($D_{z}$$>$0) and negative ($D_{z}$$<$0) DMI becomes asymmetrical in the presence of the PDI, implying the interplay of the multiple anisotropic interactions changes the topological phase diagram. 
In the individual DMI or PDI case, all the topological phases that have occurred also appear in the combined anisotropic interactions.
Particularly, several high topological-number ($C_{n}$$>$3) topological states emerge naturally in the entire topological phase diagram.
In accordance with the minimum band gap $\triangle_{Min}$, as seen in Fig.~\ref{Fig8}(b), the underlying mechanism behind the topological phase diagram can be easily explained.
It is worthy noting that the interplay of the DMI and PDI on the topological magnons in the 2D Kagome systems is distinctly different from that in the 2D honeycomb and 1D chain materials \cite{JPCM34-495801,JPCM36-255801}.

\subsection{Berry Curvature and Chern Number}
As mentioned above, the Chern number can be expressed as an integral of Berry curvature. Besides Chern number, Berry curvature can provide more detailed information about the topological state involved with non-trivial band gap. 
The anisotropic exchange interactions can induce band inversions, and open finite band gaps between the two bands.
In general, the Berry curvature is mostly polarized around the $k$ point of the band gap.   
Therefore, the topological magnon states with different Chern numbers possess completely distinct Berry curvatures.

\begin{figure}[htbp]
\hspace*{-2mm}
\centering\includegraphics[trim = 0mm 0mm 0mm 0mm, clip=true, angle=0, width=1.0 \columnwidth]{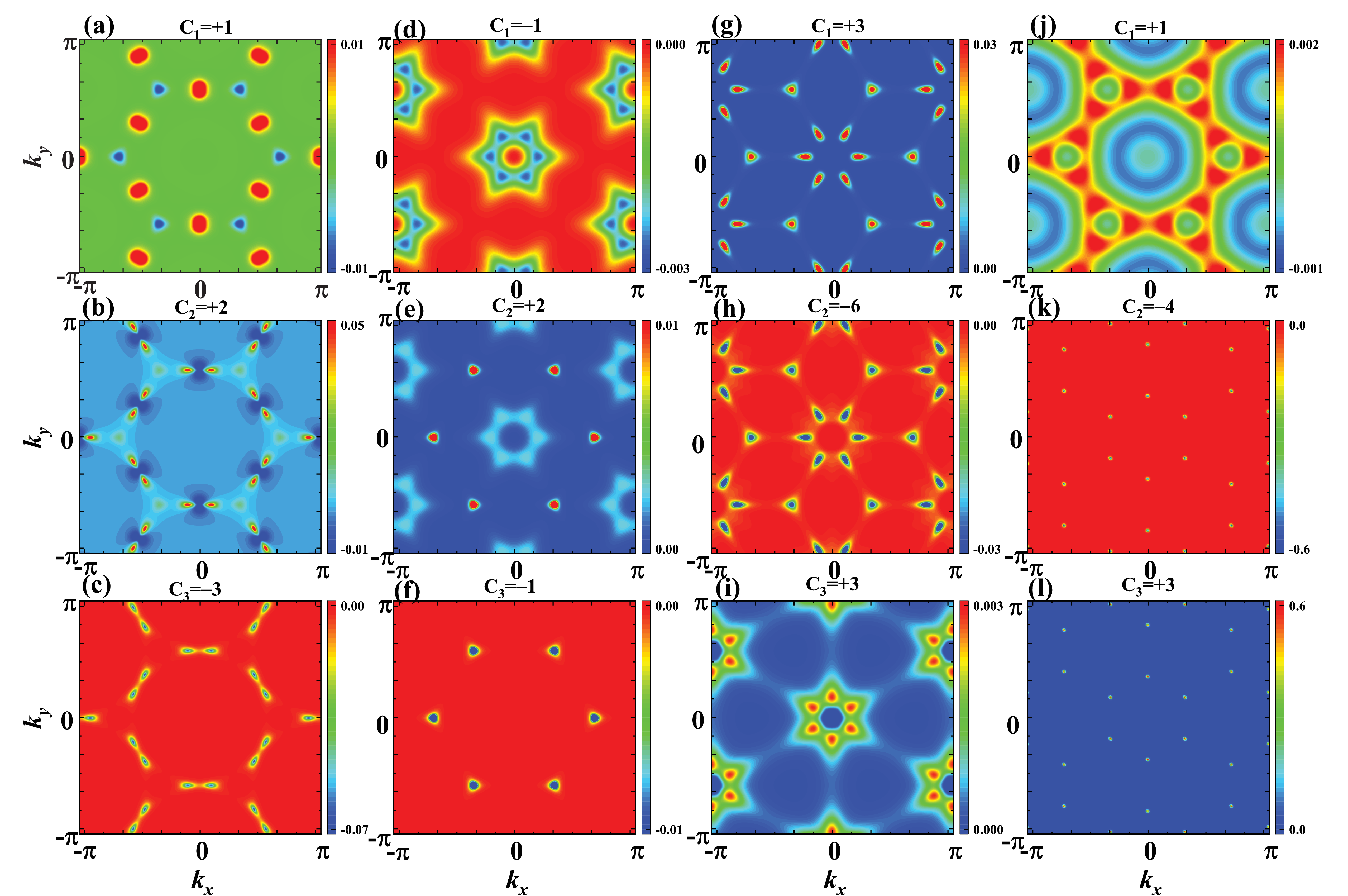}
\caption{(Color online) Berry curvatures of the upper, middle and lower bands with ($J_{2}$, $D_{z}$, $F$) being (0.8, 0.2, 0) and the corresponding $C_{n=1,2,3}$ $=$ $+$1 (a), $+$2 (b), $-$3 (c), (0.5, 0.0, 2.0) and $-$1 (d), $+$2 (e), $-$1 (f), (0.6, 0.0, $-$3.0) and $+$3 (g), $-$6 (h), $+$3 (i), (0.5, 0.2, $-$2.0) and $+$1 (j), $-$4 (k), $+$3 (l), respectively.}
\label{Fig9}
\end{figure}
Berry curvatures of the upper ($C_{1}$), middle ($C_{2}$) and lower ($C_{3}$) bands for representative topological states are displayed in Fig.~\ref{Fig9}.
Four topological states with the sets of Chern numbers ($+$1, $+$2, $-$3), ($-$1, $+$2, $-$1), ($+$3, $-$6, $+$3), and ($+$1, $-$4, $+$3), are analyzed for comparison.
As expected, the Berry curvatures of each band for different topological magnon states exhibit diverse polarization over the entire BZ.
These results also demonstrate that there exist complex band inversions between the three magnon bands, originating from the complicated interplay of multiple isotropic and anisotropic exchange interactions.  

It is known that the integral of the $n$-th band Berry curvature $B_{n}(\mathbf{k})$ over BZ is Chern number $C_{n}$, characterizing the magnon topological phase with sets ($C_{1}$, $C_{2}$, $C_{3}$) for the Kagome system.
The sum over all band Chern numbers is zero, $C_{1}+C_{2}+C_{3}=0$.
The Chern numbers with remarkably large values are found in a variety of magnon topological phases, indicating the existence of high Chern-number topological phases with $C_{n}>3$ due to the interplay of different magnetic couplings. 
The high Chern-number phases, including ($+$1, $-$4, $+$3), ($+$3, $-$4, $+$1), ($+$5, $-$4, $-$1) and ($+$3, $-$6, $+$3), can be tuned by the multiple anisotropic exchange interactions, as demonstrated in topological phase diagrams. 
Remarkably, this kind of high Chern-number phase is also found within the extended Heisenberg model with XXZ anisotropy in the Kagome lattice \cite{PRB103-054405}.

\subsection{Magnon Thermal Hall and Nernst Conductivities}
Both the DMI and PDI can induce a topological magnon insulator ({\it i.e.} magnon Chern insulator) in 2D Kagome lattice, which is characterized by the existence of one-dimensional (1D) edge states inside the non-trivial band gap.
On the other hand, the edge magnon flow is the typical feature of topological magnons, and the thermal Hall conductivity can be induced by the flow of heat current under the influence of thermal temperature gradient.
The thermal Hall conductivity of $z$ component can be written as \cite{PRB89-054420}
\begin{equation}
\kappa^{xy}=-\frac{k^{2}_{B}T}{(2\pi)^{2}\hbar}\sum\limits_{\lambda}\int_{BZ}(c_{2}(n_{\lambda})-\frac{\pi^2}{3})B_{\lambda}(\mathbf{k})d^{2}k,
\end{equation}
where $n_{\lambda}=(e^{E_{\lambda}(\mathbf{k})/{k_{B}T}}-1)^{-1}$ is the Bose-Einstein distribution function with $k_{B}$ the Boltzmann constant, and $c_{2}(x)=(1+x)(ln\frac{1+x}{x})^{2}-(lnx)^{2}-2Li_{2}(-x)$ with $Li_{2}(-x)$ the polylogarithm function. Note that $k_{B}=1$ and $\hbar=1$ in the following calculations.

\begin{figure}[htbp]
\hspace*{-2mm}
\centering\includegraphics[trim = 0mm 0mm 0mm 0mm, clip=true, angle=0, width=1.0 \columnwidth]{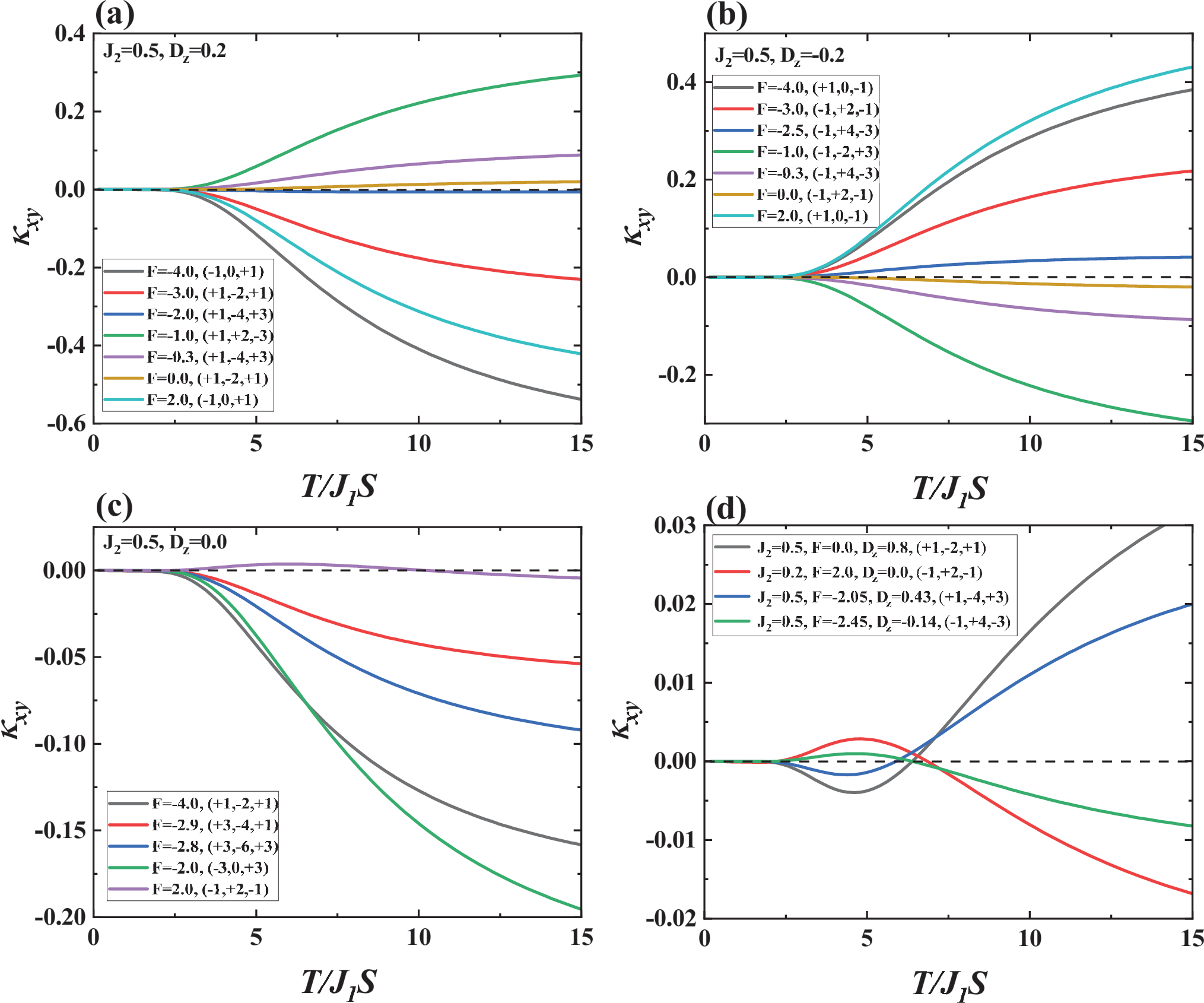}
\caption{(Color online) Thermal Hall conductivity $\kappa^{xy}$ in units of $k_{B}^2/\hbar$ as a function of temperature $T$ with parameters $J_{2}$=0.5, and (a) D$_{z}$=0.2, (b) $D_{z}$=$-$0.2, (c) $D_{z}$=0.0. (d) Sign reversals of thermal Hall conductivity for four types of topological phases.}
\label{Fig10}
\end{figure}
The calculated transverse thermal Hall conductivities $\kappa^{xy}$ as a function of temperature $T$ are shown in Fig.~\ref{Fig10}, a series of parameters are chosen for comparison.
When the parameters $J_{2}$=0.5 and $D_{z}$=0.2, $-$0.2, are fixed for comparison, the system will undergo different topological phases through varying the PDI parameter $F$ value from $-$4 to 2. And the thermal conductivity will correspondingly change, even its sign which means the direction of thermal transport can be inverted, as shown in Fig.~\ref{Fig10}(a)-(b). 
It is worth noting that the thermal Hall conductivities for the opposite DMI, {\it i.e.} $D_{z}$=0.2 and $-$0.2, is not symmetrical, owing to the presence of the PDI.
More precisely, a large $F$ value leads to more asymmetry and vice versa, which can also be seen in Fig.~\ref{Fig10}(d).
That is, the chiral DMI (sign of $D_{z}$) can induce the inverse thermal transport, corresponding to the opposite three-band Chern numbers. However, in the multiple anisotropic interactions case, the PDI can break this chirality. 
In fact, in the presence of finite $F$, no sign change of $D_{z}$ is needed to achieve sign reversal of thermal conductivity.
While for $D_{z}$=0, the topological phase transition also occurs, but there is no sign reversal of thermal Hall conductivity excluding the origin of temperature effects at $F$=2.0, as can be seen in Fig.~\ref{Fig10}(c).
Note that the sign reversal phenomena of thermal Hall conductivity contributing from the temperature effect rather than the changing of magnetic interaction parameters will be discussed in detail in the following.
\begin{figure}[htbp]
\hspace*{-2mm}
\centering\includegraphics[trim = 0mm 0mm 0mm 0mm, clip=true, angle=0, width=1.0 \columnwidth]{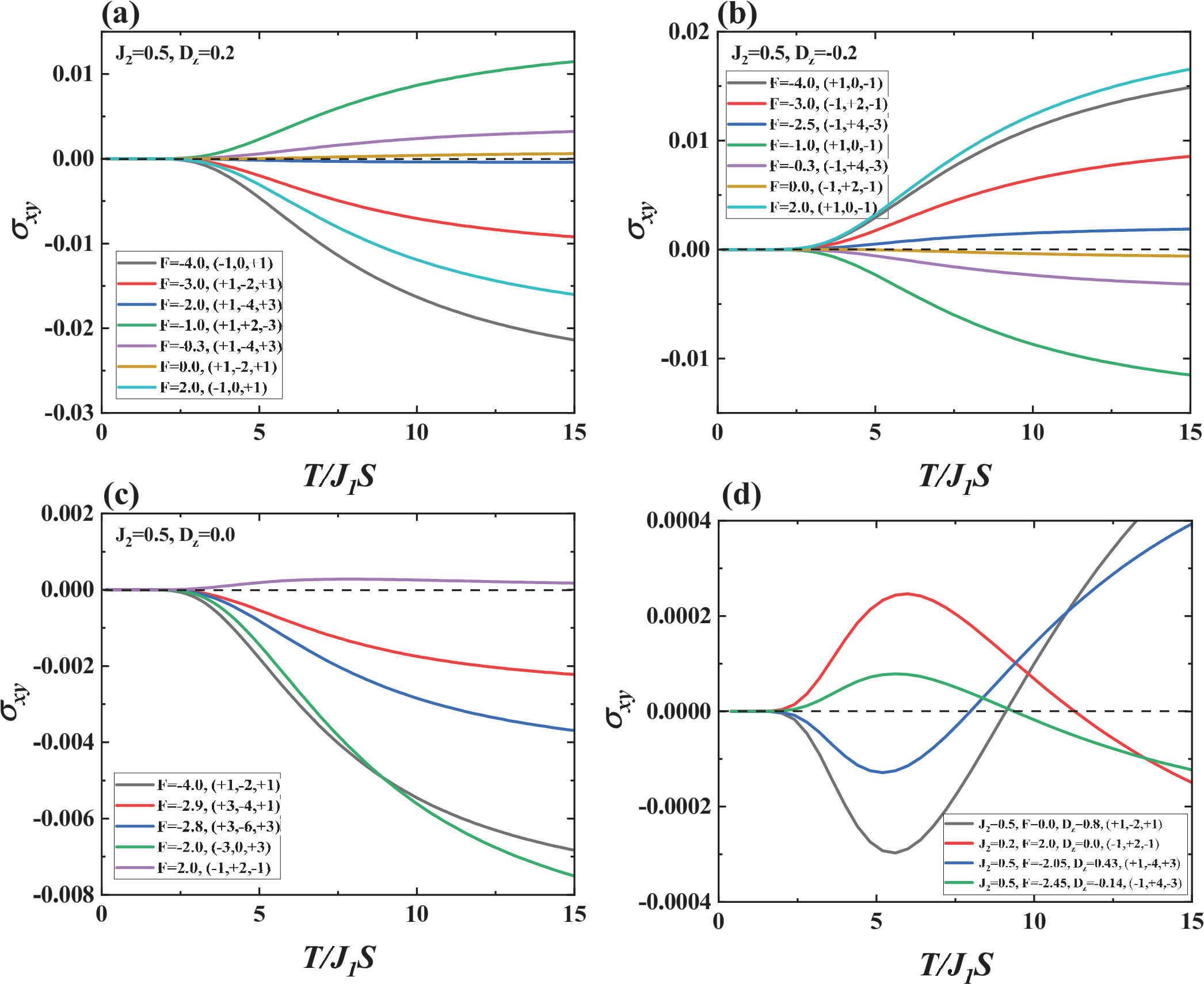}
\caption{(Color online) Magnon Nernst conductivity $\sigma^{xy}$ in units of $k_{B}$ as a function of temperature $T$ with parameters $J_{2}$=0.5, and (a) $D_{z}$=0.2, (b) $D_{z}$=$-$0.2, (c) $D_{z}$=0.0. (d) Nernst conductivity for four types of topological phases with the same parameters in Fig.~\ref{Fig10}(d).}
\label{Fig11}
\end{figure} 
In addition to the magnon thermal Hall effect, the magnon Nernst effect is another important thermo-magnetic effect in magnon system. When a temperature gradient is applied to a magnetic system, the magnons are subjected to a thermal imbalance in their distribution, which leads to a flow of magnons between different temperature regions, resulting in a magnon flow. So that a spin flow transverse to the temperature gradient occurs. The magnon Nernst conductivity $\sigma^{xy}$ can be expressed as \cite{PRB103-134414}
\begin{equation}
\sigma^{xy}=-\frac{k_{B}}{(2\pi)^{2}}\sum\limits_{\lambda}\int_{BZ}c_{1}(n_{\lambda})B_{\lambda}(\mathbf{k})d^{2}k
\end{equation}
with $c_{1}(x)=(1+x)ln(1+x)-xlnx$. Note that $k_{B}=1$ in the following calculations.
Likewise, the magnon Nernst conductivity $\sigma^{xy}$ as a function of temperature $T$ is also calculated and plotted for different parameters in Fig.~\ref{Fig11}.
Similar to the thermal Hall conductivity, the magnon Nernst conductivity also can be inversed during the topological phase transitions, as shown in Fig.~\ref{Fig11}(a) and (c). 
Especially, as displayed in Fig.~\ref{Fig11}(d), the Nernst conductivity also exhibits temperature-driven sign reversal at higher temperature, similar to the thermal Hall conductivity.

\begin{figure}[htbp]
\hspace*{-2mm}
\centering\includegraphics[trim = 0mm 0mm 0mm 0mm, clip=true, angle=0, width=0.8 \columnwidth]{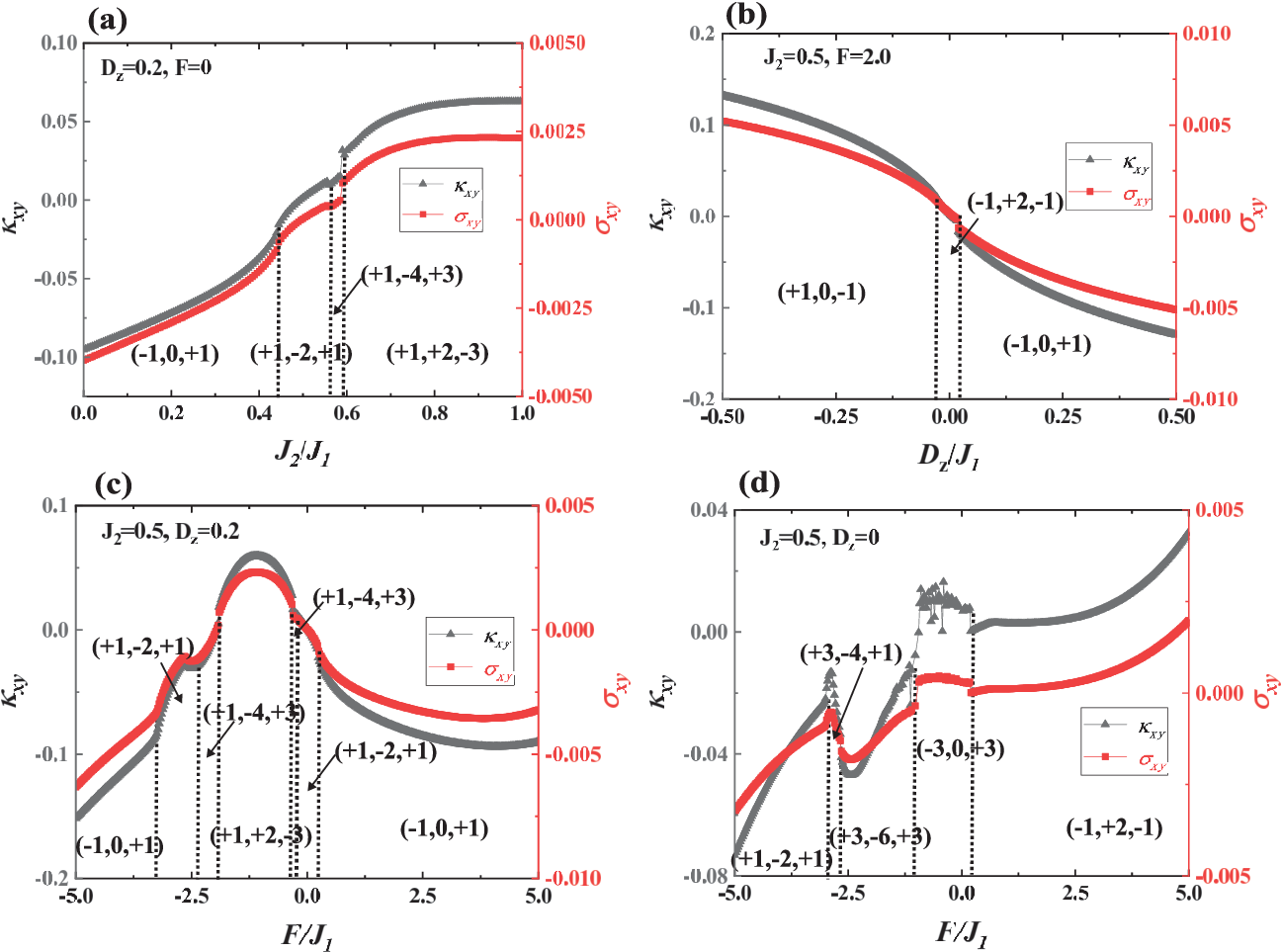}
\caption{(Color online) Magnon thermal Hall conductivity $\kappa^{xy}$ and Nernst conductivity $\sigma^{xy}$ as a function of different magnetic interaction parameters, $J_{2}$ with fixed $D_{z}$=0 and $F$=0 (a), $D_{z}$ with fixed $J_{2}$=0.5 and $F$=2.0 (b), $F$ with fixed $J_{2}$=0.5 and $D_{z}$=0.2 (c), and $F$ with fixed $J_{2}$=0.5 and $D_{z}$=0 (d), at $T=5 J_{1}S$. The dash lines indicate the distinguishable critical points of the topological phase transitions.}
\label{Fig12}
\end{figure}

Furthermore, the magnon thermal Hall and Nernst conductivities as a function of different magnetic interaction parameters, {\it i.e.} $J_{2}$, $D_{z}$ and $F$, are displayed in Fig.~\ref{Fig12}. 
The critical points of topological phase transitions can be clearly identified, since $\kappa^{xy}$ and $\sigma^{xy}$ have obviously distinguishing evolution on the magnetic interaction parameter between different magnon topological states.
Although the magnitudes of both conductivities are different, the thermal Hall and Nernst conductivities behave very similarly as a function of magnetic interaction parameters.

Now we turn to the problem of the sign change of thermal Hall and Nernst conductivities with the increase of temperature as mentioned above. The sign change means that the thermal transports flow in the opposite direction. Indeed the chiral interaction DMI ($D_{z}$) with opposite signs will result in the sign reversal of the thermal transport. In general, the thermal conductivity monotonically changes with the increase of temperature for the fixed interaction parameters. In other words, there is no sign reversal, unless the interaction parameters change and the signs of all band Chern numbers reverse.

\begin{figure}[htbp]
\hspace*{-2mm}
\centering\includegraphics[trim = 0mm 0mm 0mm 0mm, clip=true, angle=0, width=0.8 \columnwidth]{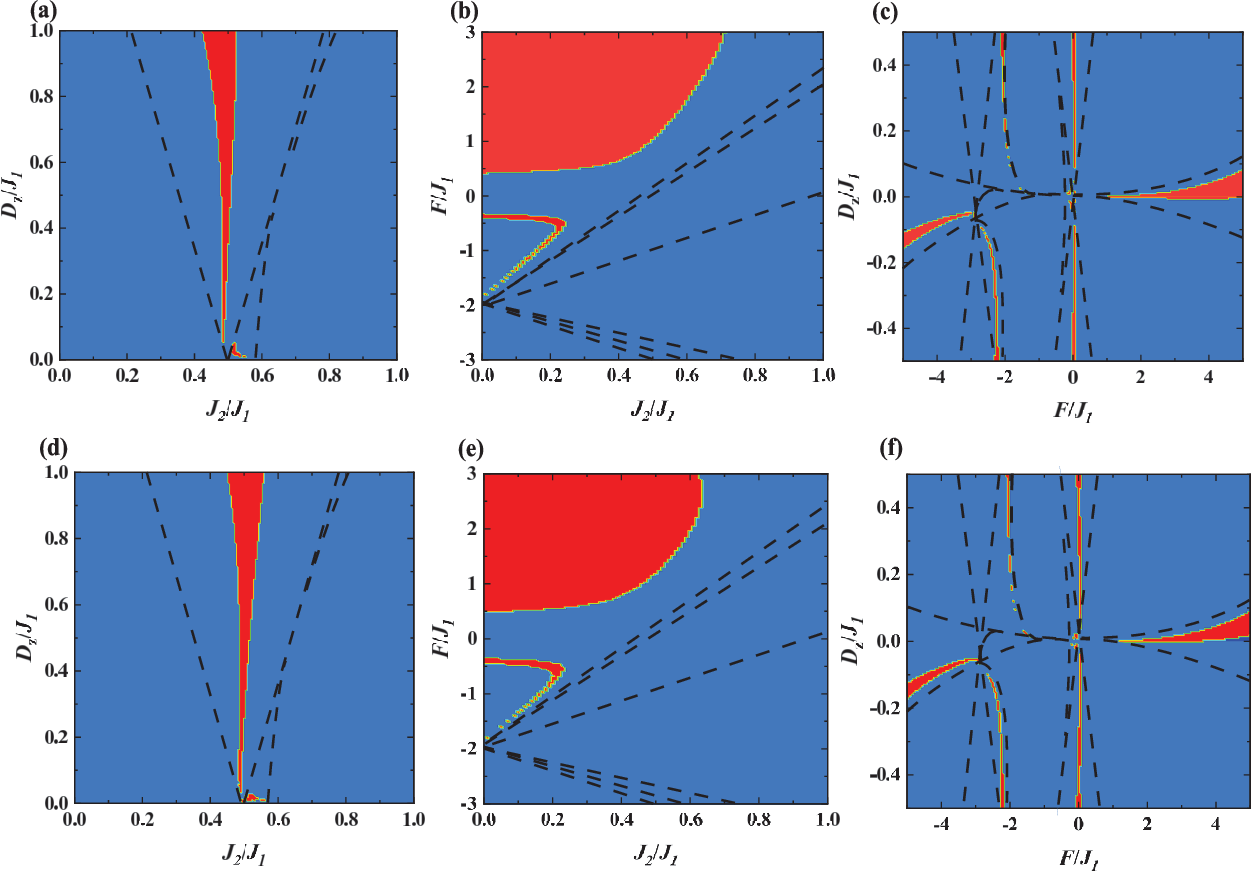}
\caption{(Color online) The $J_{2}$-$D_{z}$ (a) and (d), $J_{2}$-$F$ (b) and (e), and $F$-$D_{z}$ with fixed $J_{2}$=0.5 (c) and (f) phase diagrams for the sign reversal of thermal Hall and Nernst conductivities induced by temperature, respectively. The red (blue) region indicates the sign changed (unchanged) region. The dash lines indicate the distinguishable critical points of the topological phase transitions.}
\label{Fig13}
\end{figure}

Unlike sign reversal phenomena that stem from magnetic interactions, the temperature-induced sign reversal of the thermal Hall and Nernst conductivities still remains unclear.
Interestingly, the sign reversal phenomenon of thermal Hall conductivity induced by thermal fluctuations are also observed in honeycomb Kitaev magnet $\alpha$-RuCl$_{3}$ \cite{PRL120-217205,PRB99-085136,arXiv2001.01899,PRB103-174402}, and Na$_{2}$Co$_{2}$TeO$_{6}$ \cite{arXiv2201.11396}.
In addition, a series of sign reversals of $\kappa^{xy}$ depending on the temperature in honeycomb magnets had also been realized by tuning magnetic field and electrical field experimentally \cite{PhysRep1070-1} and theoretically \cite{PRB110-045423,arXiv2410.10355}.
Recently, this anomalous thermal Hall transport has been experimentally reported in 2D Kagome insulator volborthite Cu$_{3}$V$_{2}$O$_{7}$(OH)$_{2}$$\cdot$2H$_{2}$O magnets \cite{PNAS113-8653}. 
Interestingly, the sign reversal of the Nernst conductivity $\sigma^{xy}$ is also observed experimentally in honeycombic MnPS$_{3}$ \cite{PRB96-134425}.
Moreover, the sign reversal phenomena are extensively explored by the experimental \cite{PRL115-106603,PhysRep1070-1} and theoretical \cite{PRB98-094419,PRB91-125413,NJP24-023033} researches in Kagome AFM and FM systems.
To explore its origins, the thermal Hall and Nernst conductivities $\kappa^{xy}$ and $\sigma^{xy}$ calculations are revisited for the entire phase diagrams. Through exploring the signs of $\kappa^{xy}$ and $\sigma^{xy}$ as a function of temperature $T$, the sign reversal regions colored red are determined in three typical phase diagrams, which are displayed in Fig.~\ref{Fig13}.
Considering the effective magnitude of the numerical accuracy and the experimental observability, the criteria for the sign-reversal phase diagrams of the thermal Hall and Nernst conductivities $\kappa^{xy}$ and $\sigma^{xy}$ are set to 0.001 and 0.0001, respectively. Values below the criteria are deemed insignificant and can be disregarded.
It is worth noting that the phase diagrams for the sign reversal of thermal Hall and Nernst conductivities are so similar that both should probably have the same origin. 
It can be very clearly found that the sign reversal is not an accidental phenomenon, but a common one for a certain range of magnetic interaction parameters in topological phase diagrams.
More importantly, it usually occurs in special temperature range for different magnetic interaction parameters.
It is demonstrated that the sign reversal is linked to the band gaps and Berry curvatures (Chern numbers) in the vicinity of magnetic phase transition owing to the thermal fluctuations.
It was surprising to find that all the sign-reversal regions fall into two topological phases, {\it i.e.} ($-$1, $+$2, $-$1)/($+$1, $-$2, $+$1) and ($-$1, $+$4, $-$3)/($+$1, $-$4, $+$3), in three typical topological phase diagrams.
Although the origin of the sign reversal of magnon thermal Hall  conductivity is very complicated, it can be approximately estimated in the high temperature limit according to previous studies \cite{JPCM27-166003,PRB89-134409,PRB103-134414} for simplicity.
Assuming that without band crossing, $\kappa_{lim}^{xy} \varpropto -\sum_{n}C_{n}\varepsilon_{n}$, with $\varepsilon_{1}>\varepsilon_{2}>\varepsilon_{3}$.
Considering that $C_{1}\varepsilon_{1}+C_{2}\varepsilon_{2}+C_{3}\varepsilon_{3}
=C_{1}(\varepsilon_{1}-\varepsilon_{2})-C_{3}(\varepsilon_{2}-\varepsilon_{3})=C_{1}\Delta_{1,2}-C_{3}\Delta_{2,3}$,
the sign of $\kappa_{lim}^{xy}$ is unique with the opposite sign for $C_{1}$ and $C_{3}$. 
By contraries, the sign of $\kappa_{lim}^{xy}$ can be possibly changed within the topological phases with the same sign of $C_{1}$ and $C_{3}$, such as ($-$1, $+$2, $-$1)/($+$1, $-$2, $+$1), ($-$1, $+$4, $-$3)/($+$1, $-$4, $+$3), ($-$3, $+$4, $-$1)/($+$3, $-$4, $+$1), and ($-$3, $+$6, $-$3)/($+$3, $-$6, $+$3).
As expected, we did find the sign reversal phenomena in ($-/+$1, $+/-$2, $-/+$1) and ($-/+$1, $+/-$4, $-/+$3). While it is absent in the other two topological phases ($-/+$3, $+/-$4, $-/+$1) and ($-/+$3, $+/-$6, $-/+$3) due to the narrow region in the present phase diagrams.
The sign-reversal phase diagrams of the Nernst and thermal Hall conductivities should be roughly consistent, except near the boundary of the phase diagram.
Indeed, the sign-reversal of conductivity is determined by the characteristic band structure including energy spectrum, band gap, and competition between the middle band and the upper and lower bands, and its topological properties.
Our results reveal the topological origin of the sign reversal of the magnon thermal Hall and Nernst conductivities.
Consequently, this topological origin scenario provides a possible explanation for the experimental puzzles \cite{PRL115-106603,PNAS113-8653,PhysRep1070-1}.

\section{Remarks and Conclusions}
The multiple topological phases in the topological phase diagrams in this study can be realized in extensive transition-metal Kagome ferromagnetic systems discovered in experiments \cite{npjCM6-158,PRB106-115139,arXiv2412.02010,PRL125-217202,JPCM29-493002,PRB89-014414,PRB98-134437, PRL115-147201,PNAS113-8653,NC15-1592,SCIENCE329-297}, and the strong magnetic anisotropy can be realistically achievable in Kagome Ising ferromagnets and Kagome Kitaev systems \cite{npjCM6-158,PRB106-115139,arXiv2412.02010,JPCM29-493002,PRB89-014414,PRB98-134437}. The most promising candidates for the Kagome magnets with strong magnetic anisotropy are the experimentally observed Kagome Ising ferromagnet TbV$_{6}$Sn$_{6}$ \cite{PRB106-115139,arXiv2412.02010}, and the theoretically predicted Kagome Kitaev iridates Na$_{4}$Ir$_{3}$O$_{8}$ and Ir$_{2}$O$_{4}$ \cite{JPCM29-493002,PRB89-014414}, and the rare-earth-based Kagome Kitaev compound A$_{2}$RE$_{3}$Sb$_{3}$O$_{14}$ \cite{PRB98-134437}.

In summary, the topological magnons in a two-dimensional Kagome ferromagnet are investigated based on a Heisenberg model together with multiple anisotropic exchange interactions including the Dzyaloshinskii-Moriya and pseudo-dipolar ones.
Distinctly different from the Dzyaloshinskii-Moriya and Kitaev types of interactions, the pseudo-dipolar interaction naturally induce multiple topological magnons including high Chern-number non-trivial states, and complex topological phase diagrams owing to the coexistence of the Dirac and flat bands in Kagome lattice.
Moreover, the interplay between the multiple anisotropic exchange interactions provides diverse topological phase transitions due to the band inversion and gap closing-reopening \cite{JPCM34-495801,JPCM36-255801}, implying that both cooperation and competition between the Dzyaloshinskii-Moriya and pseudo-dipolar interactions play an essential role.
In addition, a topological origin of the sign reversal of the thermal Hall and Nernst conductivities induced by temperature is proposed to explain the experimental puzzles.
To realize the spin model used here, the 3-5$d$ and 4-5$f$ correlated materials with both spin-orbit coupling and orbital localized states, such as iridates and ruthenates, can achieve the required magnetic properties accordingly.
Further experimental and theoretical work is desired to explore the regulation of topological magnon states and their transport behaviors in realistic magnetic materials with multiple anisotropic exchange interactions.
Our results reveal the lavish topological magnons in Kagome magnets, indicating the tunable topological regulations and potential applications in topological quantum devices.

\acknowledgements
We acknowledge support from the National Natural Science Foundation of China (Grant Nos. 11974354, 11774350, and 11574315). Numerical calculations are performed at the Center for Computational Science of CASHIPS.

\end{document}